\documentclass[preprints,article,accept,moreauthors,pdftex]{Definitions/mdpi}
\usepackage{changes}

\newcommand{\declarecommand}[1]{\providecommand{#1}{}\renewcommand{#1}}
\declarecommand{\q}{\mathbf{q}}
\declarecommand{\vk}{\mathbf{k}}
\declarecommand{\p}{\mathbf{p}}
\declarecommand{\Q}{\mathbf{Q}}
\declarecommand{\eps}{\varepsilon}
\declarecommand{\su}{\uparrow}
\declarecommand{\sd}{\downarrow}
\declarecommand{\sgn}[1] {\mathrm{sgn}\left({#1}\right)}
\declarecommand{\sgnnobr}[1] {\mathrm{sgn}{#1}}
\declarecommand{\sign}{\mathrm{sgn}}
\declarecommand{\abs}[1] {\left|{#1}\right|}
\declarecommand{\ii}{{\mathrm{i}}}
\declarecommand{\av}[1]{\left<{#1}\right>}
\declarecommand{\bra}[1]{\left<{#1}\right|}
\declarecommand{\ket}[1]{\left|{#1}\right>}

\firstpage{1}
\pubvolume{1}
\issuenum{1}
\articlenumber{0}
\pubyear{2022}
\copyrightyear{2022}
\datereceived{}
\dateaccepted{}
\datepublished{}
\hreflink{https://doi.org/} 

\Title{One- and two-particle correlation functions in the cluster perturbation theory for cuprates}

\TitleCitation{One- and two-particle correlation functions in the CPT for cuprates}

\Author{V.I. Kuz'min $^{1}$\orcidA{}, S.V. Nikolaev $^{1,2}$\orcidB{}, M.M. Korshunov$^{1,2,}$*\orcidC{}, and S.G. Ovchinnikov $^{1,2}$\orcidD{}}

\AuthorNames{V.I. Kuz'min, S.V. Nikolaev, M.M. Korshunov, and S.G. Ovchinnikov}

\AuthorCitation{Kuz'min V. I.; Nikolaev, S. V.; Korshunov, M. M.; Ovchinnikov S. G.}

\address{%
$^{1}$ \quad Kirensky Institute of Physics, Federal Research Center KSC SB RAS, Akademgorodok, 660036, Krasnoyarsk, Russia\\
$^{2}$ \quad Siberian Federal University, Svobodny Prospect 79, 660041 Krasnoyarsk, Russia}

\corres{Correspondence: mkor@iph.krasn.ru}

\abstract{Physics of high-$T_c$ superconducting cuprates is obscured by the effect of strong electronic correlations. One way to overcome the problem is to seek for an exact solution at least within the small cluster and expand it to the whole crystal. Such an approach is in the heart of the cluster perturbation theory (CPT). Here we develop CPT for the dynamic spin and charge susceptibilities (spin-CPT and charge-CPT), within which the correlation effects are explicitly taken into account by the exact diagonalization. We apply spin-CPT and charge-CPT to the effective two-band Hubbard model for the cuprates obtained from the three-band Emery model and calculate one- and two-particle correlation functions, namely, spectral function and spin and charge susceptibilities. Doping dependence of the spin susceptibility was studied within spin-CPT and CPT-RPA that is the CPT generalization of the random phase approximation (RPA). Both methods produce the low energy response at four incommensurate wave vectors in qualitative agreement to the results of the inelastic neutron scattering on overdoped cuprates.}

\keyword{Strong electronic correlations; Cuprates; Hubbard model; Cluster perturbation theory}

\PACS{74.25.-q, 74.45.+c, 74.70.Xa, 74.20.Fg}

\begin{document}

\section{\label{Intro} Introduction}

Electronic and magnetic subsystems of a correlated material are strongly coupled. Classic example is the cuprate high-$T_c$ superconductors, whose phase diagram contains antiferromagnetic (AFM), metallic, charge-density wave, and pseudogap states. Undoped cuprates corresponding to the half-filled system is the AFM insulator. In this case the spin dynamics corresponds to the excitations of magnons and could be described by the linear spin-wave theory. The physics, however, becomes more complicated once the holes are doped into the system. Holes are delocalized and the AFM long-range order quickly disappears opening the way for the new cooperative phases to arise. Recent resonant inelastic X-ray scattering (RIXS) reveal magnon-like excitations in cuprates over a wide doping range from underdoped to heavily overdoped systems~\cite{Letacon11,Letacon13,Wang22,Robarts19}. Moreover, the momentum-dependent charge excitations are observed in RIXS in the pseudogap phase of cuprates~\cite{Ishii17}. Spin and charge dynamics is coupled to the doping-dependent changes of the electronic structure observed in angle-resolved photoemission spectroscopy (ARPES)~\cite{Sobota_2021}. While electronic excitations can be described by the one-particle Green's functions, calculation of the magnetic and charge excitations rely on the two-particle response functions. Two-particle correlation functions provide important information about ordered phases of a strongly correlated system. On the experimental side, inelastic neutron scattering allows to observe the dynamical spin susceptibility, two-particle spin-spin correlation function, directly~\cite{Tranquada14,Fujita12,Sato20,Lipscombe09,Hinkov07,Chan16,Chan16prl}.

Cluster perturbation theory (CPT) is a straightforward way of calculating the one-particle correlation function, i.e. the spectral function~\cite{Senechal00,Senechal02}. Latter can be directly compared to ARPES data. CPT is one of a number of quantum cluster theories~\cite{Maier2005}, also, it is the most economical cluster method in terms of the necessary computation power. In CPT, the first step is the exact diagonalization (ED) of a small cluster. Therefore, the short-range correlations are treated exactly. At the second step, the intercluster interactions are included according to some kind of a perturbation theory. There have been few attempts to expand the CPT for the calculation of two-particle correlation functions. The authors of Ref.~\cite{Brehm_2010} used the variational cluster approximation (modified CPT by a self-consistent procedure) to solve the Bethe-Salpeter equation for a two-dimensional Hubbard model. In Ref.~\cite{Kung17}, the spin susceptibility is calculated within the determinant quantum Monte Carlo method and CPT for the one-band Hubbard model. The authors of Ref.~\cite{Raum20} extended the CPT to compute the two-particle correlation functions by approximately solving the Bethe-Salpeter equation for the one-band one-dimensional Hubbard model.

Cuprates have a quasi-two-dimensional structure and the conductivity is provided by the electrons in the copper-oxygen plane. This was the reason for intensive studies of two-dimensional models and, in particular, effective low-energy Hubbard model as a simple one yet retaining essential physics. However, a detailed study of the electronic and magnetic properties of cuprates requires a more realistic model such as the three-band Emery model that includes both Cu-$d_{x^2-y^2}$ and O-$p_{x,y}$ orbitals~\cite{Emery}.

Here we develop the CPT for the dynamical spin susceptibility, the approach we call \textit{spin-CPT}. It is based on an explicit calculation of correlation functions by the exact diagonalization with a subsequent extraction of the lattice two-particle spectral function in the CPT-like manner similar to Refs.~\cite{Chen15,Kung17,Parschke19}. Then we apply spin-CPT to the two-dimensional model of cuprates -- effective Hubbard model for CuO plane based on the Emery model. We compare the results of the spin-CPT and the CPT-RPA approach~\cite{Nikolaev16,Raum20}. The latter is a straightforward generalization of the random phase approximation (RPA), where the bare electron Green's functions forming the electron-hole bubble are replaced by the ones obtained within CPT. We show that spin-CPT produces the low-energy response at four incommensurate wave vectors that qualitatively agrees with the results of the inelastic neutron scattering on overdoped cuprates. Also, it allows to obtain a spectral intensity distribution resembling the upper branch of an experimentally observed hourglass dispersion in a wide doping range.

This paper is organized as follows. In Section~\ref{Model} we discuss the model and approximations used for the study. In Section~\ref{Results} the main results are presented. In Section~\ref{Discussion} the results are discussed. In Section~\ref{Conclusion} the concluding remarks are given.

\section{\label{Model} Model and methods}

\subsection{Model}

The Hamiltonian of the Emery model~\cite{Emery} describes the copper $d_{x^2-y^2}$ and oxygen $p_{x,y}$ orbitals,
\begin{eqnarray}
 H_{pd} &=& \sum_{i,\sigma} \varepsilon_d n^d_{i\sigma} + \sum_{j,\sigma} \varepsilon_p n^p_{j\sigma} + \lambda U_d \sum_{i} n^d_{i\uparrow}n^d_{i\downarrow} + \lambda U_p \sum_{j} n^p_{j\uparrow}n^p_{j\downarrow} + \lambda V_{pd} \sum_{\left<i,j\right>} n^d_{i}n^p_{j} \nonumber \\
 &+& t_{pd} \sum_{\left<i,j\right>,\sigma} \left(-1\right)^{P_{ij}} \left(d^{\dag}_{i\sigma}p^{}_{j\sigma} + \text{H.c.}\right) + t_{pp} \sum_{\left<\left< j,j'\right>\right>,\sigma} \left(-1\right)^{P'_{jj'}} \left(p^{\dag}_{j\sigma}p^{}_{j'\sigma} + \text{H.c.}\right),
\label{eq:1}
\end{eqnarray}
where $i$ denotes copper sites, $j$ denotes oxygen sites, $n_{i(j)\sigma}$ is the number operator of holes on a site $i(j)$ with spin $\sigma$, $n_{i(j)} = n_{i(j)\uparrow} + n_{i(j)\downarrow}$, $P_{ij}$ and $P'_{jj'}$ are phase factors, and $d^{}_{i\sigma}$ ($p^{}_{j\sigma}$) destroys a hole with spin $\sigma$ on a $d_{x^2-y^2}$ ($p_{x(y)}$) orbital and site $i$ ($j$). The Coulomb interaction normalization constant $\lambda$ is introduced for convenience to be able to vary the interaction strength.

The following hopping parameters are used (here and below all energies are in eV): $t_{pd} = 1.36$, $t_{pp} = 0.86$. These are the hopping integrals of multiband $p-d$ model calculated for La$_2$CuO$_4$~\cite{Korshunov05} by projecting~\cite{Anisimov05} the \textit{ab initio} local density approximation electronic structure onto the Wannier function basis. The Coulomb parameters used here are $U_d = 9$, $U_p = 4$, and $V_{pd} = 1.5$. The charge-transfer energy parameter is taken to have a typical value, $\varepsilon_p = 3.6$~\cite{Hybertsen89}. The one-electron energy of the copper $d$-orbital is set to zero, $\varepsilon_d = 0$.

Since CPT has a cluster at it's heart, choice of it is important. In a CuO plane, one can naturally distinguish a single element consisting of a copper ion surrounded by four oxygens. Oxygen, however, belongs to the two neighboring cells. To overcome this difficulty, we use the canonical fermions of Shastry~\cite{Shastry} as done in Refs.~\cite{Lovtsov91,Jefferson92,Schuttler92,Feiner96,Raimondi96,Gavrichkov00,Makarov15,Shneyder20}, thereby orthogonalizing $p_{x,y}$ orbitals and ending up with a representation of a square lattice of orthogonal CuO$_4$ cells,
\begin{eqnarray}
 H &=& \sum_{f,\sigma} \left[\sum_{\alpha}\varepsilon_{\alpha} n^{\alpha}_{f\sigma} - 2t_{pd}\mu_{00}\left(d^{\dag}_{f\sigma}b^{}_{f\sigma} + b^{\dag}_{f\sigma}d^{}_{f\sigma} \right)\right] + \sum_{f, \alpha} \lambda U_{\alpha} n^{\alpha}_{f\uparrow}n^{\alpha}_{f\downarrow} + \sum_{f, \alpha < \beta} \lambda V_{\alpha\beta} n^{\alpha}_{f}n^{\beta}_{f} \nonumber \\
 &+& \sum_{f\neq g, \sigma} \left[-2t_{pd}\mu_{fg}\left(d^{\dag}_{f\sigma}b^{}_{g\sigma} + b^{\dag}_{f\sigma}d^{}_{g\sigma} \right) + 2t_{pp}\nu_{fg} \left(a^{\dag}_{f\sigma}a^{}_{g\sigma} - b^{\dag}_{f\sigma}b^{}_{g\sigma}\right) \right. \nonumber \\
 &-& \left. 2t_{pp}\chi_{fg}\left(a^{\dag}_{f\sigma}b^{}_{g\sigma} + b^{\dag}_{f\sigma}a^{}_{g\sigma} \right) \right] + H^{cc}_{int},
\label{eq:2}
\end{eqnarray}
where $f$ indexes a cell position (site), $\varepsilon_{\alpha}$ is a one-electron energy for an orbital index $\alpha$, which takes values $\left\{a, b, d\right\}$, where $d$ is the copper orbital and $a$ and $b$ are the cell oxygen orbitals with the energies $\varepsilon_a = \varepsilon_p + 2t_{pp} \nu_{00}$ and $\varepsilon_b = \varepsilon_p - 2t_{pp} \nu_{00}$. The operator $d^{}_{f\sigma}$ annihilates a hole with spin $\sigma$ on a copper orbital on a site $f$, operators $a^{}_{f\sigma}$ ($b^{}_{f\sigma}$) annihilate a hole with spin $\sigma$ on an oxygen orbital $a$ ($b$) on a site $f$, $n^{\alpha}_{f\sigma}$ are the number operators, $n^{\alpha}_{f} = n^{\alpha}_{f\uparrow} + n^{\alpha}_{f\downarrow}$. The Wannier coefficients $\mu$, $\nu$ and $\chi$ and other constants relating parameters of Equations~(\ref{eq:1}) and (\ref{eq:2}) are the same as in Refs.~\cite{Feiner96,Gavrichkov00,Makarov15,Shneyder20}. In particular, the effective Coulomb interactions and the original ones are related as $U_a = U_b \approx 0.21 U_p$, $V_{ad} = V_{bd} \approx 0.91 V_{pd}, U_{ab} \approx 0.17 U_p$. Finally, $H^{cc}_{int}$ includes all non-local interaction terms including three-site and four-site ones, which do not have a significant influence on the results obtained using Equation~(\ref{eq:2}), and it is omitted hereafter.

\begin{figure}
\centering
\includegraphics[width=1.0\linewidth]{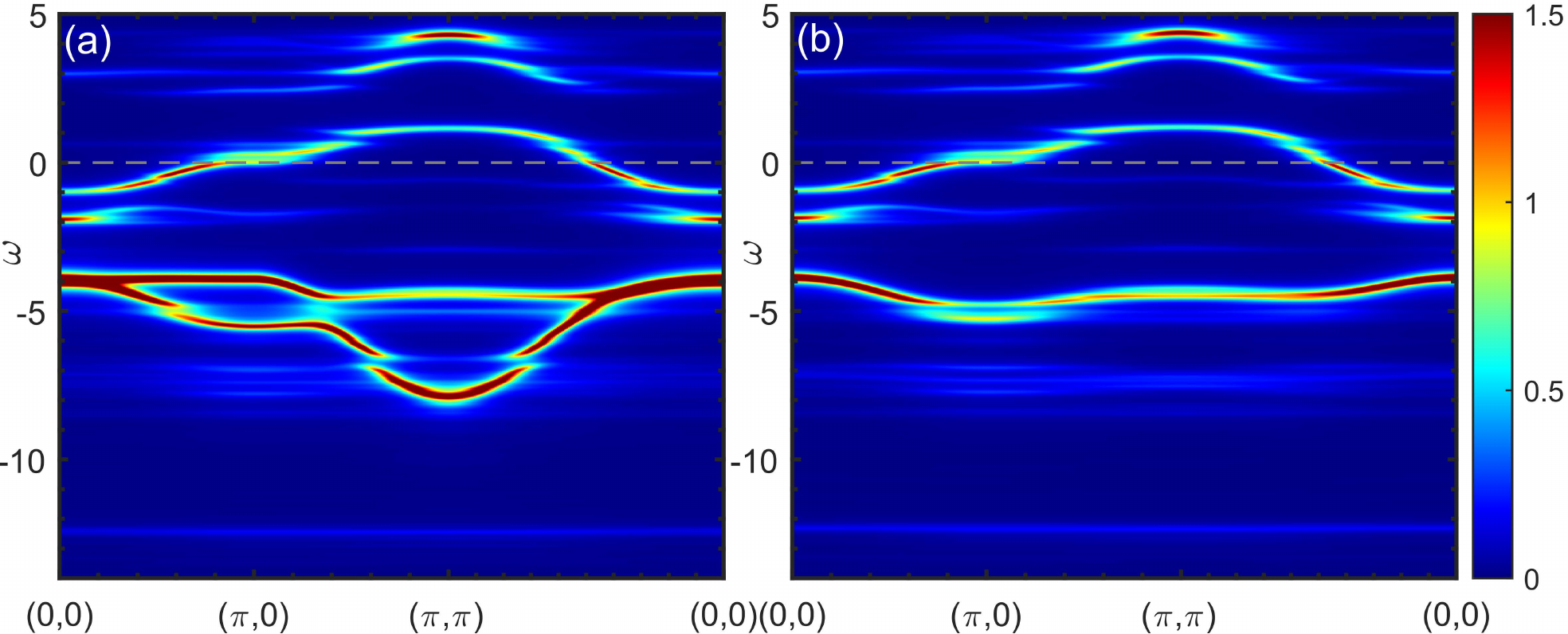}
\caption{\label{Fig_compare_A} The electronic spectral function calculated within CPT using a $2 \times 2$ cluster for the Hamiltonian~(\ref{eq:2}) including (a) and excluding (b) the $a$ orbital. A Lorentzian broadening $\eta=0.1$ is used here and below.}
\end{figure}

Such a cell representation is especially useful when one is interested in the low-energy properties. Since only $b$ orbitals are directly connected by the hopping processes with $d$ orbitals, they should have a negligible effect on the low-energy part of the electronic structure. On the other hand, at high doping $a$ orbitals might become important~\cite{Jefferson92}. In Figure~\ref{Fig_compare_A}, the comparison of the spectral functions obtained with and without the $a$ orbital is shown for the overdoped case with doping $p=0.25$ using a $2 \times 2$ cluster consisting of four orthogonal cells. The distinguishable difference is seen only 2 eV below the Fermi level within the bands dominated by the oxygen spectral weight~\cite{Kung16}. Since the needed computational power rises with the number of possible states that increase with the number of orbitals considered, we can increase the cluster size once we remove one of the orbitals. Thus, in the main part of the paper we omit the $a$ orbitals. This allows us to extend the cluster size within CPT to be a $3 \times 3$ consisting of nine orthogonal cells with the moderate computational effort.

\subsection{Methods}

The central object we are going to calculate is the transverse part of the dynamical spin susceptibility $\chi_{\alpha\beta, \gamma\delta}(\mathbf{q},\Omega)$. It can be defined as an analytical continuation of the Matsubara two-particle spin-spin correlation function
\begin{equation}
 \chi_{\alpha\beta, \alpha'\beta'}(\mathbf{q},\ii\omega_n) = \int\limits_0^{1/T} d\tau e^{\ii\omega_n\tau} \av{T_\tau S^{+}_{\alpha\alpha'}(\mathbf{q},\tau) S^{-}_{\beta'\beta}(-\mathbf{q},0)}.
\label{eq.chi_def}
\end{equation}
where $S^{+}_{\alpha\alpha'}$ and $S^{-}_{\beta'\beta}$ are the spin raising and lowering operators, $\mathbf{q}$ is the wave vector, $\omega_n$ is the Matsubara frequency, $\alpha$, $\beta$, $\alpha'$, $\beta'$ are the orbital indices, $T_\tau$ is the ordering operator over Matsubara (imaginary) time $\tau$, and $T$ is the temperature in the energy units. Averaging is done with the fully interacting Hamiltonian. By using the Wick's theorem, one can obtain the zeroth order approximation for the spin susceptibility in terms of the one-particle Green's function $G_{\alpha\beta \sigma}(\mathbf{p},\ii\omega_m)$ for an electron with the spin $\sigma$,
\begin{equation}
 \chi_{\alpha\beta, \alpha'\beta'}^{(0)}(\mathbf{q},\ii\omega_n) = -T \sum_{\omega_m, \mathbf{p}} G_{\alpha\beta \su}(\mathbf{p}, \ii\omega_m) G_{\alpha' \beta' \sd}(\mathbf{p}+\mathbf{q}, \ii\omega_n + \ii\omega_m).
\label{eq.chipm}
\end{equation}
Analytical continuation $\ii\omega_n \to \Omega + \ii\eta$ includes positive infinitesimal parameter $\eta$ that can be used to introduce a Lorentzian broadening in the course of the numerical calculations. Since we consider the paramagnetic phase, the spin index will be omitted.

The physical spin susceptibility is calculated on real frequencies $\Omega$ as the sum over all orbital indices coinciding at vertices,
\begin{equation}
 \chi(\mathbf{q},\Omega) = \dfrac{1}{2} \sum_{\alpha, \beta}{\chi_{\alpha\alpha,\beta\beta}(\mathbf{q},\Omega)}.
\label{eq.chi_phys}
\end{equation}

Analogously to the spin correlation function~(\ref{eq.chi_def}), the charge correlation function is defined as
\begin{equation}
 \Pi_{\alpha\beta, \alpha'\beta'}(\mathbf{q},\ii\omega_n) = \int\limits_0^{1/T} d\tau e^{\ii\omega_n\tau} \av{T_\tau \rho_{\alpha\alpha'}(\mathbf{q},\tau) \rho_{\beta'\beta}(-\mathbf{q},0)}.
\label{eq.Pi_def}
\end{equation}
where $\rho_{\alpha\alpha'}$ is the particle density operator. The physical charge susceptibility on real frequencies $\Omega$ is calculated as
\begin{equation}
 \Pi(\mathbf{q},\Omega) = \dfrac{1}{2} \sum_{\alpha, \beta}{\Pi_{\alpha\alpha,\beta\beta}(\mathbf{q},\Omega)}.
\label{eq.Pi_phys}
\end{equation}

\subsubsection{CPT}

Electronic structure of the two-band model is studied within CPT that is discussed in details in Refs.~\cite{Senechal00,Senechal02}. There are two steps in this approach. First one is the exact solution for the small cluster usually achieved by the exact diagonalization of the model Hamiltonian. Thus, the lattice is partitioned into a superlattice of clusters with a new translational order of an artificial origin. Second step is the reconstruction of the whole lattice by adding intercluster interactions that is done in a perturbative manner.

For the Hamiltonian~(\ref{eq:2}), we calculate the electronic spectral function
\begin{equation}
 A_{\alpha\beta}(\mathbf{k},\omega) = -\dfrac{1}{\pi} \mathrm{Im} G_{\alpha\alpha}(\mathbf{k},\omega + \ii\eta),
 \label{eq:A}
\end{equation}
where $G_{\alpha\beta}(\mathbf{k},\omega)$ is the electronic Green's function on the real frequencies $\omega$, $\mathbf{k}$ is a wave vector, $\alpha$ and $\beta$ are the orbital indices, and $\eta \to 0^+$. Trace of~(\ref{eq:A}) gives the total (physical) spectral function $A(\mathbf{k},\omega) = \sum\limits_{\alpha}{A_{\alpha\alpha}(\mathbf{q}, \omega)}$. CPT allows to calculate $A(\mathbf{k},\omega)$ within a small cluster and an arbitrary momentum resolution; it is also characterized by a fast convergence of results with increasing cluster size~\cite{Huang21}. Here, the intracluster interactions are taken into account by means of ED of a cluster with open boundary conditions using the Lanczos method, while the intercluster hoppings are treated perturbatively. This approach allows to include the short-range (intracluster) correlations explicitly. The calculations are done at zero temperature using a square cluster consisting of nine cells (sites).

The main CPT equation is
\begin{equation}
 \hat{G}(\mathbf{\tilde{k}},\omega)^{-1} = \hat{G}^c(\omega)^{-1} - \hat{T}(\mathbf{\tilde{k}}),
\label{eq:3}
\end{equation}
where all square matrices have dimension $L \times N_c$, $L$ is the number of orbitals that's in our case equal to two, $N_c$ is the number of sites within the cluster, $\mathbf{\tilde{k}}$ is a wave vector defined within the cluster Brillouin zone, $\omega$ is a frequency, $\hat{G}^c(\omega)$ is the exact local (intracluster) propagator, and $\hat{T}(\mathbf{\tilde{k}})$ is a Fourier transform of the hopping matrix. The translational invariance is restored in CPT as
\begin{equation}
 G_{\alpha\beta}(\mathbf{k},\omega) = \frac{1}{N_c} \sum_{i, j} e^{-i\left(\mathbf{r}_i - \mathbf{r}_j\right)\mathbf{k}} {G_{\alpha\beta, i j}(\mathbf{k},\omega)},
\label{eq:4}
\end{equation}
where $i(j)$ is an intracluster site index, $\mathbf{r}_{i(j)}$ is a corresponding radius-vector.

\subsubsection{CPT-RPA}

CPT can be used to improve calculation of the dynamical spin susceptibility $\chi(\mathbf{k},\Omega)$ within RPA by replacing the bare electron-hole bubble, $\chi^{(0)}(\mathbf{k},\Omega)$, by the one calculated using CPT Green's functions. It allows to include self-energy corrections over the vertex renormalization of RPA. The methodology is discussed in details in Refs.~\cite{Nikolaev16,Raum20}. Thus, given an RPA vertex, the obtained magnetic structure is determined by the electronic spectrum. Instead of the ``bare'' susceptibility, we have the electron-hole bubble calculated via the cluster electronic spectral functions~\cite{Nikolaev16}:
\begin{equation}
 \chi_{\alpha\beta, \gamma\delta}^{c(0)}(\mathbf{k},\Omega) = - \sum_{\mathbf{q}}\int\int d\omega' d\omega'' A_{\alpha\beta}(\mathbf{q}, \omega') A_{\gamma\delta}(\mathbf{k} + \mathbf{q}, \omega'') \dfrac{f(\omega') - f(\omega'')}{\omega' - \omega'' + \Omega + \ii\eta},
\label{eq:chi0}
\end{equation}
where $f(\omega)$ is the Fermi function, and $\eta$ is the positive infinitesimal.

The transverse dynamical susceptibility within CPT-RPA is calculated as~\cite{Graser09,Korshunov_chapter}
\begin{equation}
 \chi_{\alpha\beta, \gamma\delta}(\mathbf{k},\Omega) = \chi_{\alpha\beta, \gamma\delta}^{c(0)}(\mathbf{k},\Omega) + \sum_{\alpha', \beta', \gamma', \delta'} \chi_{\alpha\beta, \alpha'\beta'}^{c(0)}(\mathbf{k},\Omega) \bar{U}^{\alpha'\beta'}_{\gamma'\delta'} \chi_{\gamma'\delta', \gamma\delta}(\mathbf{k},\Omega).
\label{eq:5}
\end{equation}
A vertex $\mathbf{\bar{U}}$ including another renormalization constant $\lambda'$ that will be discussed below is defined as $\mathbf{\bar{U}}=\lambda'\mathbf{U}$, where the nonzero components of the RPA vertex $\mathbf{U}$ are $U^{\alpha\alpha}_{\alpha\alpha} = U_\alpha$ and $U^{\alpha\beta}_{\alpha\beta} = V_{\alpha\beta}$, where $\alpha \neq \beta$ \cite{Graser09, Korshunov_chapter}. The particle-hole bubble $\chi^{c0}(\mathbf{k},\Omega)$ is calculated in this paper using the CPT spectral function as in Ref.~\cite{Nikolaev16}.

Dynamical charge susceptibility $\Pi(\mathbf{q},\Omega)$ is obtained within the CPT-RPA by replacing the vertex $\mathbf{U}$ in the RPA equation~(\ref{eq:5}) with the vertex $-\mathbf{V}$, where $V^{\alpha\alpha}_{\alpha\alpha} = U_{\alpha}$, $V^{\alpha\alpha}_{\beta\beta} = 2V_{\alpha\beta}$, $V^{\alpha\beta}_{\alpha\beta} = -V_{\alpha\beta}$~\cite{Graser09}.

\subsubsection{Explicit calculation of the two-particle correlation function (spin-CPT and charge-CPT)}

Alternative to the RPA-like approach is to calculate the two-particle correlation function directly without intermediate steps of finding the one-particle Green's function and building a spin or charge correlation function out of it. Here we calculate the right side of Equation~(\ref{eq.chi_def}) in a CPT-like manner and refer to this approach as \textit{spin-CPT}. One-particle CPT takes the intracluser correlations exactly and treats the intercluster ones approximately, for example, within the Hubbard-I approximation for the Green's functions built on the Fermi-type Hubbard operators~\cite{Nikolaev10,Nikolaev12,Kuzmin14,Kuzmin20}. A straightforward application of this ideology with respect to Bose-type Hubbard quasiparticles forming the spin excitations leads to an approximation in which intercluster interaction does not enter the Green's function calculation procedure since no spin-spin interaction terms to be taken into account by the generalized Hubbard-I approximation are present in Hamiltonian~(\ref{eq:2}). Moreover, the intercluster vertex corrections have been shown to not producing a crucial effect in such cluster calculations~\cite{Chen15,Kung17,Parschke19}. Thus, the whole procedure is given by two steps and it is identical to the implemented in Refs.~\cite{Kung17,Parschke19}. At first, the transverse spin susceptibility is calculated by the ED within the cluster via the following equation
\begin{equation}
 \chi^c_{\alpha\beta, ij}(\Omega) = \sum_\mu\left[\frac{\left< 0 \right| S^+_{\alpha,i} \left|\mu\right> \left< \mu \right| S^-_{\beta,j} \left| 0 \right> }{\Omega - \left(E_\mu - E_0\right) + \ii\eta} - \frac{\left< 0 \right| S^-_{\beta,j} \left|\mu\right> \left< \mu \right| S^+_{\alpha,i} \left| 0 \right> }{\Omega + \left(E_\mu - E_0\right) + \ii\eta}\right],
\label{eq:chi_ED}
\end{equation}
where $S^{-}$ and $S^{+}$ are the spin ladder operators, $\left|0\right>$ and $\left|\mu\right>$ are the ground and excited states, which are obtained in the Lanczos approximations. Then the translational invariance for the whole lattice is restored analogously to Equation~(\ref{eq:4}):
\begin{equation}
 \chi_{\alpha\beta}(\mathbf{k},\Omega) = \frac{1}{N_c} \sum_{i, j} e^{-i\left(\mathbf{r}_i - \mathbf{r}_j\right)\mathbf{k}} {\chi^c_{\alpha\beta, ij}(\Omega)}.
\label{eq:chi_spinCPT}
\end{equation}

Dynamical charge susceptibility $\Pi(\mathbf{q},\Omega)$ can be also calculated in a similar manner on a cluster as
\begin{equation}
 \Pi^c_{\alpha,\beta,i,j}(\Omega) = \sum_\mu\left[\frac{\left< 0 \right| n_{\alpha,i} \left|\mu\right> \left< \mu \right| n_{\beta,j} \left| 0 \right> }{\Omega - \left(E_\mu - E_0\right) + \ii\eta} - \frac{\left< 0 \right| n_{\beta,j} \left|\mu\right> \left< \mu \right| n_{\alpha,i} \left| 0 \right> }{\Omega + \left(E_\mu - E_0\right) + \ii\eta}\right],
\label{eq:Pi_ED}
\end{equation}
where $n_{\alpha,i}$ is the number of particle operator. Since such an approach allows to calculate the charge correlation function directly, we refer to it as \textit{charge-CPT}.

\begin{figure}
\centering
\includegraphics[width=1.0\linewidth]{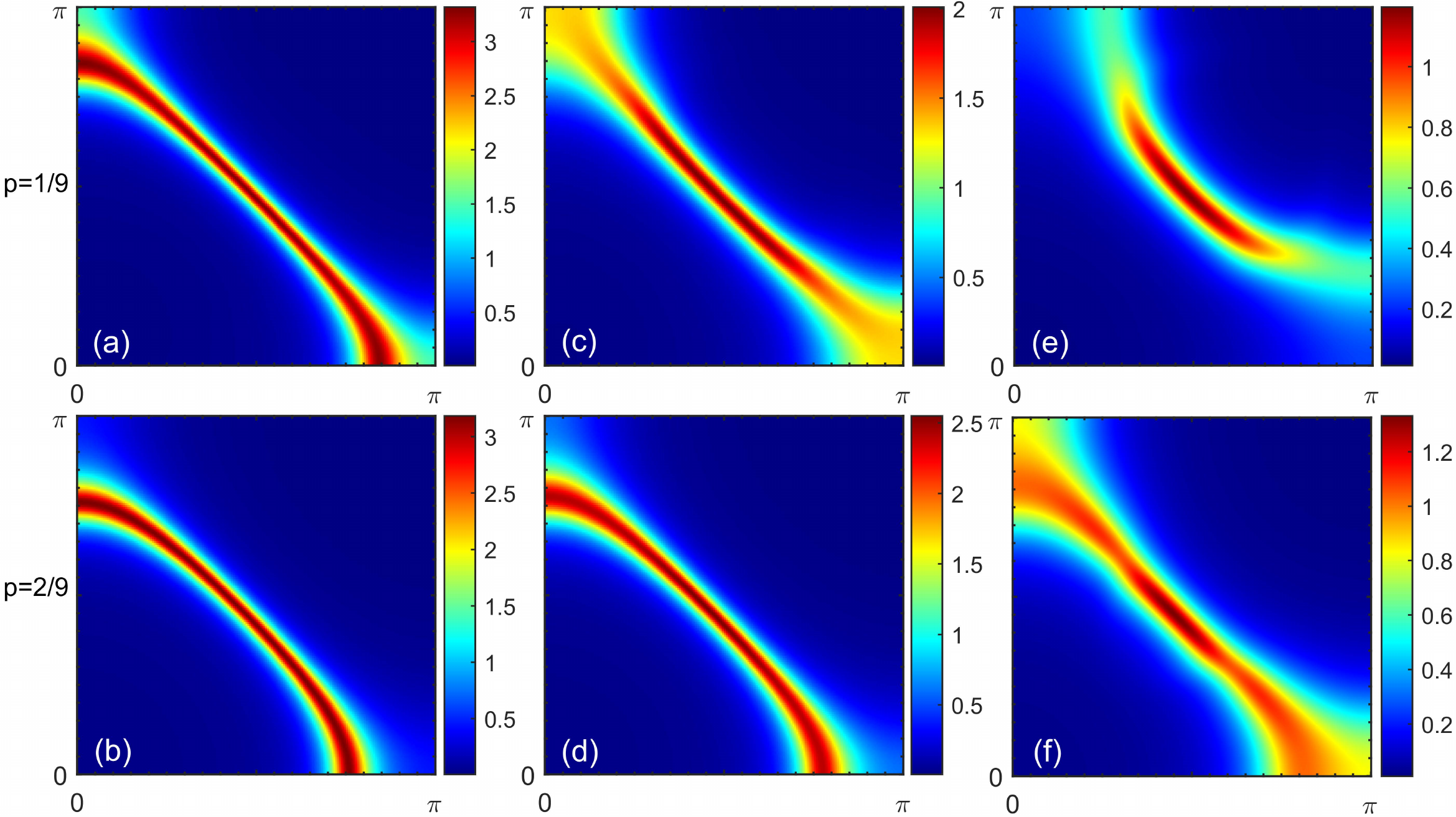}
\caption{\label{Fig_Fermi} The electronic spectral function at the Fermi level for doping $p=1/9$ (a), (c), (e) and $p=2/9$ (b), (d), (f) obtained  using the Coulomb interaction normalization constant $\lambda=0$ (a), (b), $\lambda=0.3$ (c), (d), and $\lambda=1$ (e), (f).}
\end{figure}

\section{\label{Results} Results}

\begin{figure}
\centering
\includegraphics[width=1.0\linewidth]{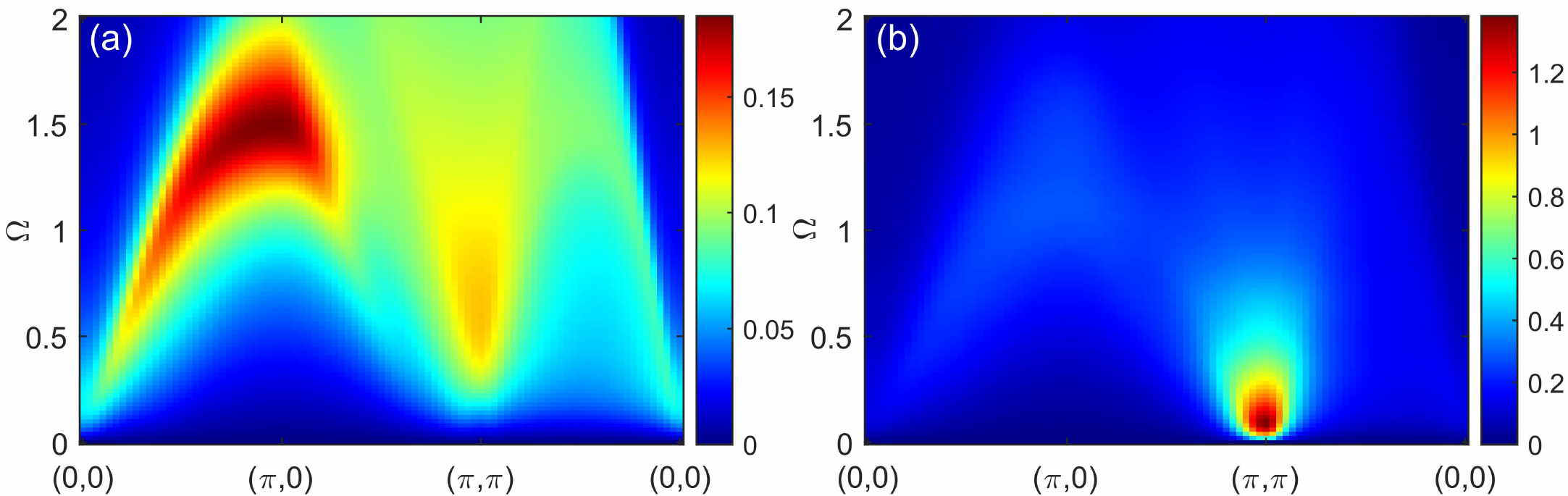}
\caption{\label{Fig_hi_U} Dynamical spin susceptibility $\chi(\mathbf{q},\Omega)$ calculated within CPT-RPA with a $3\times3$ cluster for doping $p=1/9$ and Coulomb interaction renormalization constants $\lambda=\lambda'=0$ (a) and $\lambda=\lambda'=0.3$ (b).}
\end{figure}

To see the changes of the low-energy electronic structure with doping and interactions, we start with the calculation of the Fermi surface for the CuO model~(\ref{eq:2}). Two doping values corresponding to an integer number of electrons per cluster are shown in Figure~\ref{Fig_Fermi} where we present the spectral function at the Fermi level thus giving the idea of how the Fermi contour would look like in ARPES. The noninteracting Fermi surface is electron-like. Increase of the Coulomb interaction by changing the parameter $\lambda$ affects the Fermi surface dramatically at low doping. For $\lambda=0.3$, it is already hole-like with the spectral weight decreasing from the nodal to the antinodal direction that is a signature of the pseudogap. On the contrary, the correlation effect is negligible at large doping. Lifshitz transition occurs in between the presented values of doping at $p \approx 0.13$. For the interaction of a full strength, $\lambda=1$, the pseudogap is clearly visible at low doping while the electron-like Fermi surface at $p=2/9$ affected only slightly. This time, the change of topology appears at $p \approx 0.17$.

\begin{figure}
\centering
\includegraphics[width=1.0\linewidth]{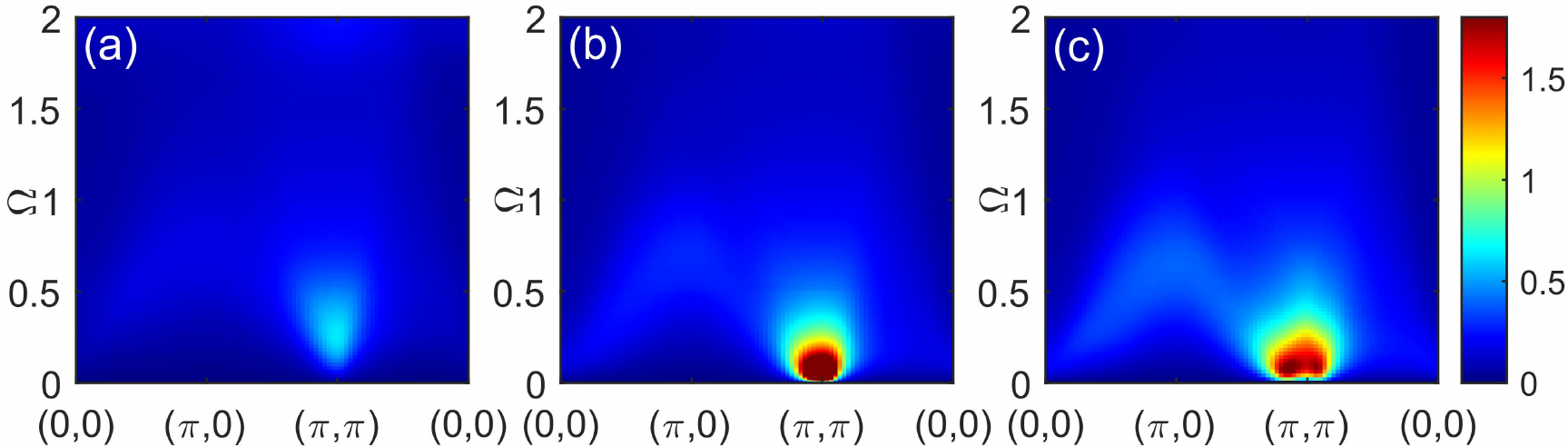}
\caption{\label{Fig_hi_p} Dynamical spin susceptibility $\chi(\mathbf{q},\Omega)$ calculated within CPT-RPA with $\lambda=1$ and $\lambda'=0.75$ for dopings $p=1/9$ (a), $p=1/6$ (b), and $p=2/9$ (c).}
\end{figure}

Now we turn to the CPT-RPA results, which demonstrate how the spin spectrum is affected by the one-particle processes entering the particle-hole bubble $\chi^{c0}(\mathbf{k},\Omega)$ through the spectral functions and enhanced by the RPA vertex. Dynamical spin susceptibility within CPT-RPA with a Lorentzian broadening $\eta = 0.1$ are presented in Figures~\ref{Fig_hi_U} and~\ref{Fig_hi_p}. At low doping and relatively weak interaction strength (Figure~\ref{Fig_hi_U}), changes in electronic structure like the emergence of the pseudogap result in the formation of a response resembling a spin wave spectrum. In the lowest-energy region, it is dominated by a contribution of the spectral weight at the antiferromagnetic wave vector due to the presence of the short-range antiferromagnetic correlations dominating the electronic spectrum.

\begin{figure}
\centering
\includegraphics[width=1.0\linewidth]{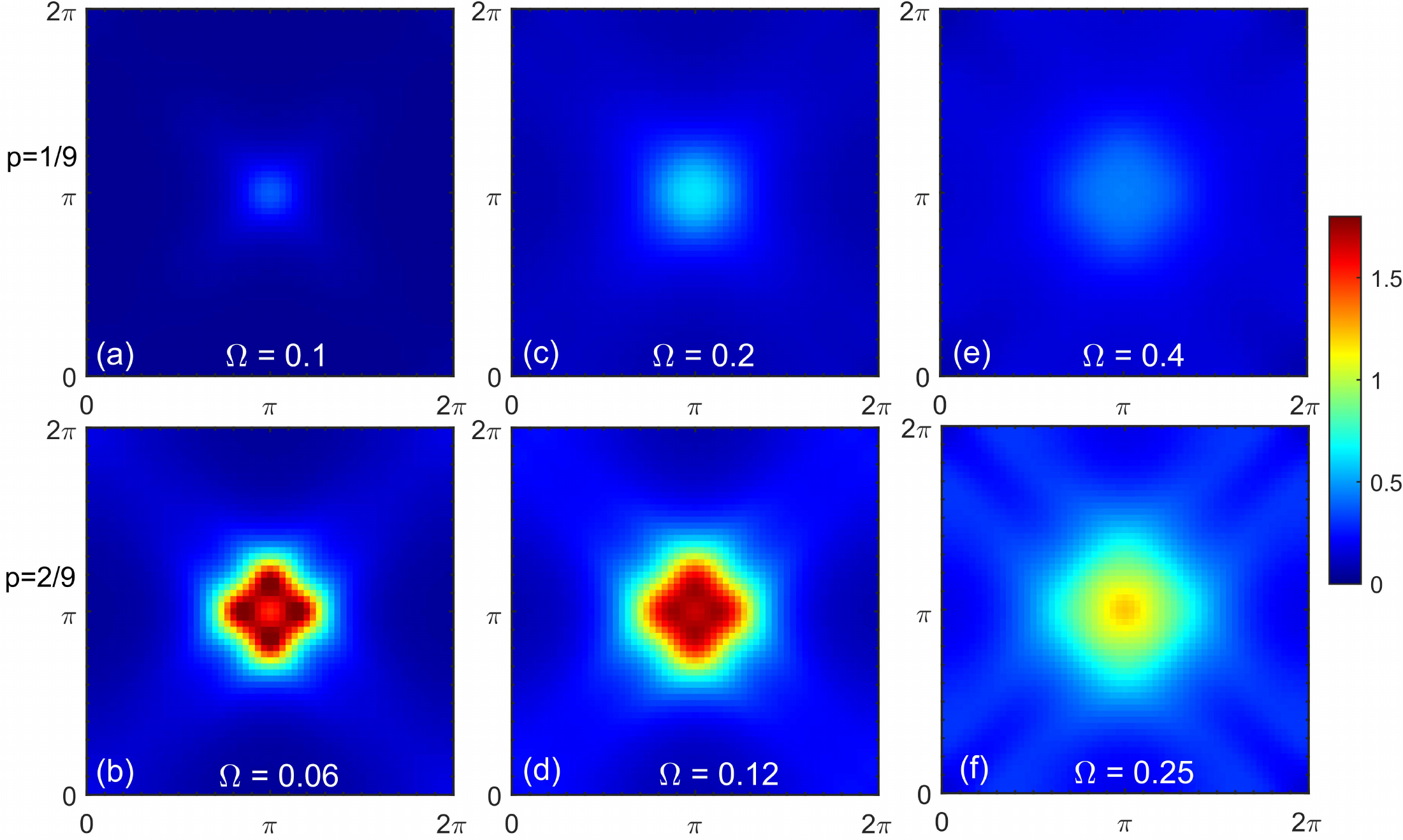}
\caption{\label{Fig_hi_iso} Constant energy cuts of the dynamical spin susceptibility calculated within CPT-RPA with $\lambda=1$ and $\lambda'=0.75$ for dopings $p=1/9$ (a), (c), (e) and $p=2/9$ (b), (d), (f) at energies $\Omega$ (corresponding value is shown in each panel).
}
\end{figure}

Figure~\ref{Fig_hi_p} shows the doping dependence of the magnetic spectrum within CPT-RPA at strong interaction. Here, the Coulomb interaction entering the CPT and $\chi^{c0}(\mathbf{k},\Omega)$ is at the full strength with $\lambda=1$. Parameter $\lambda'$ that controls the strength of the interactions entering the RPA vertex, however, is taken to be less than unity to avoid the magnetic instability at finite doping, i.e. the divergence of the real part of zero-frequency RPA spin susceptibility within CPT-RPA. We adjust $\lambda'$ so that the system is in the vicinity of an instability, which appears first at the Lifshitz transition, where the particle density at the Fermi level is maximal ($p \approx 0.17$ in the present case). In the pseudogap regime, the response is again somehow similar to the spin wave spectrum, however, it is highly damped. The energy of the $(\pi,\pi)$ spectral weight maximum tends to zero at the Lifshitz transition point, see Figure~\ref{Fig_hi_p}(b). A qualitative change with doping occurs in the overdoped case, where the low-energy spectral weight peaks shift to the incommensurate wave vectors, see Figure~\ref{Fig_hi_p}(c). It is clearly seen in the corresponding constant-energy slices shown in Figure~\ref{Fig_hi_iso}. At small doping, the maximal spectral weight is confined near the antiferromagnetic wave vector over the wide energy range. At high doping, on the contrary, low-energy excitations emerge at incommensurate wave vectors forming a cross-like shape with the maximal spectral weight in the antinodal direction. Increase of the excitations energy leads to a decrease in the incommensurability.

\begin{figure}
\centering
\includegraphics[width=1.0\linewidth]{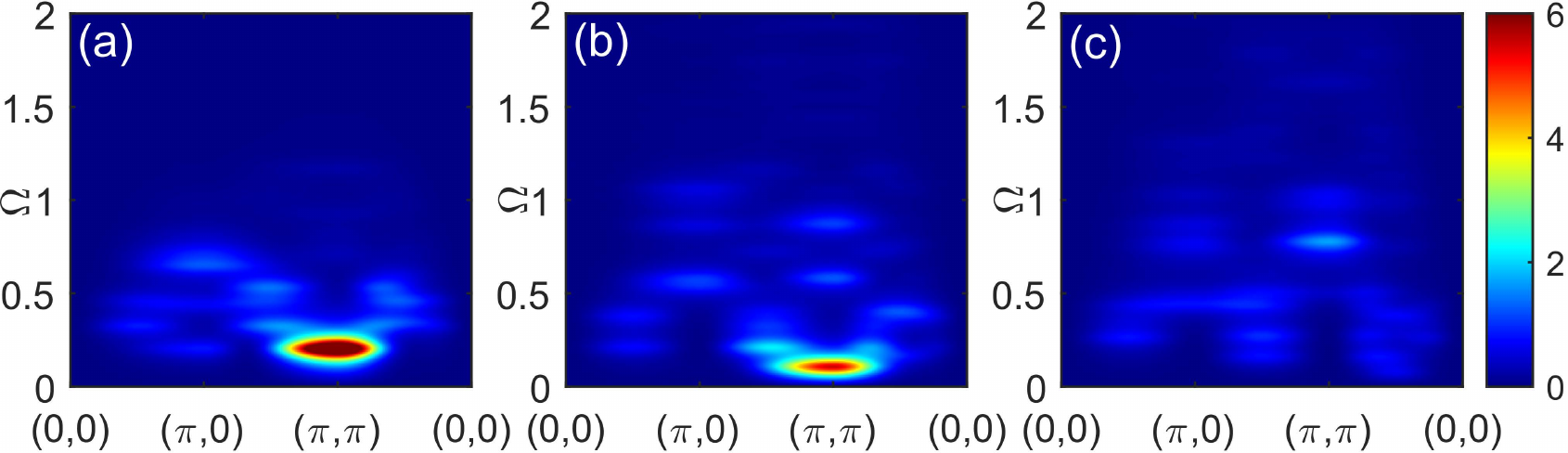}
\caption{\label{Fig_hi_SCPT} Dynamical spin susceptibility $\chi(\mathbf{q},\Omega)$ calculated within spin-CPT for dopings $p=0$ (a), $p=1/9$ (b), and $p=2/9$ (c).}
\end{figure}

\begin{figure}
\centering
\includegraphics[width=1.0\linewidth]{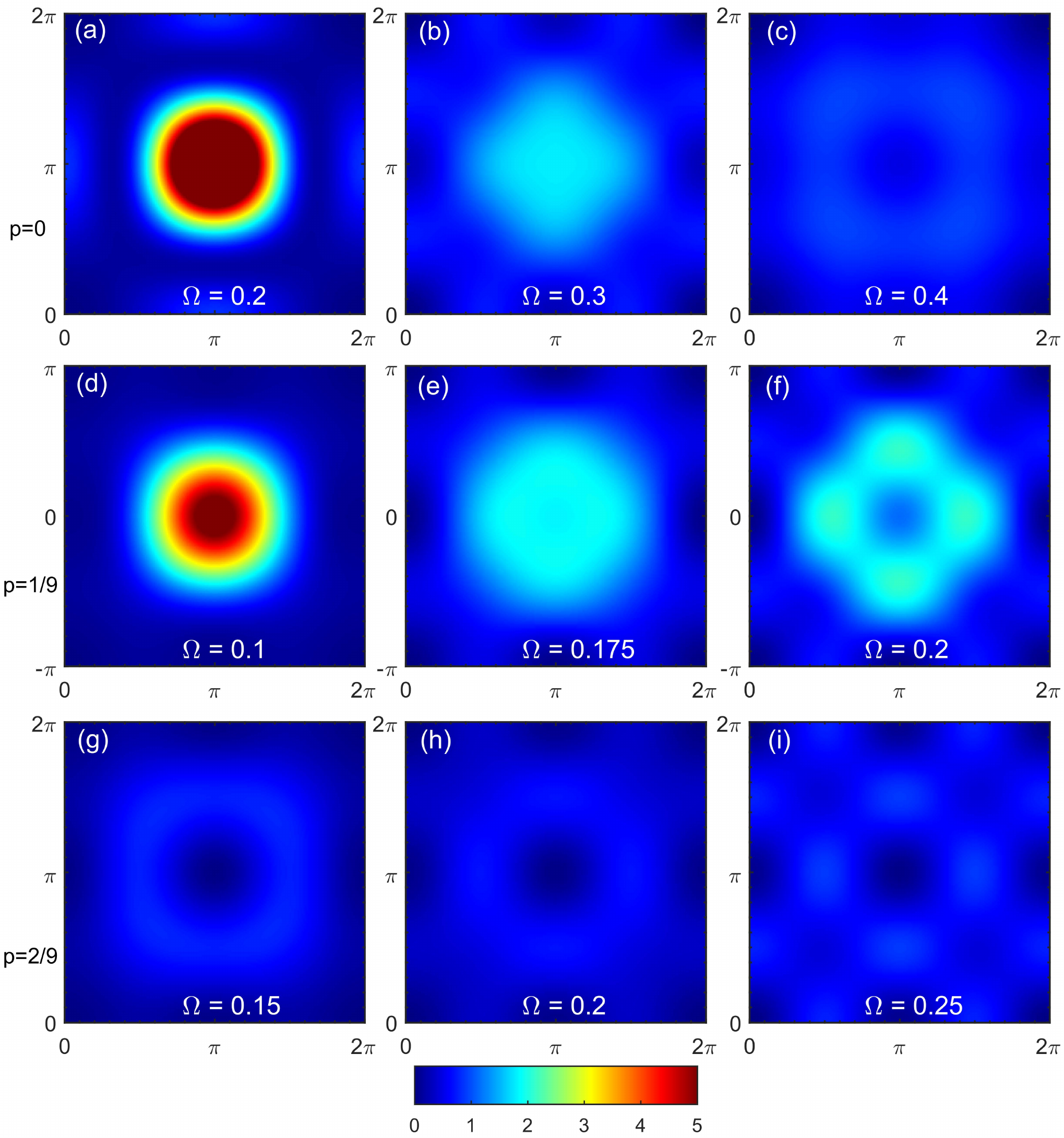}
\caption{\label{Fig_hi_iso2} Constant energy cuts of the dynamical spin susceptibility within spin-CPT for dopings $p=0$ (a), (b), (c), $p=1/9$ (d), (e), (f), and $p=2/9$ (g), (h), (i) at energies $\Omega$ (corresponding value is shown in each panel).}
\end{figure}

The doping dependence of the spin susceptibility within another approach, spin-CPT, where correlation effects are treated explicitly by ED is shown in Figure~\ref{Fig_hi_SCPT}. At half filling, the result is substantially different from picture in the CPT-RPA. The response function in Figure~\ref{Fig_hi_SCPT}(a) clearly resembles a spin wave spectrum. However, since only short-range correlations are present within a cluster, a maximal intensity at the antiferromagnetic wave vector is shifted to higher frequencies compared to the case of a long-range order. The corresponding constant-energy cuts are shown in Figure~\ref{Fig_hi_iso2} with values of energies chosen to demonstrate the most prominent features for each doping. At higher energies, the cuts reveal a dip near the commensurate wave vector $(\pi,\pi)$ expected for damped spin-waves and not seen in CPT-RPA. At a small doping, the results are quite similar but the spin response is less sharply peaked at $(\pi,\pi)$. The energy evolution of constant-energy cuts in Figures~\ref{Fig_hi_iso2}(b), (f), and (i) shows the formation of a cross-like feature at higher energies with the peaks of the spectral weight shifted from the center to the antinodal direction. In the overdoped regime with $p=2/9$, the spectral peak at $(\pi,\pi)$ is seen only at high frequency in Figure~\ref{Fig_hi_SCPT}(c). All low-energy response is at incommensurate wave vectors and is more pronounced in the antinodal direction, see Figure~\ref{Fig_hi_iso2}(i).

Note that some typical artifacts of the cluster calculations such as size effects are present in the results. Those effects should not be taken as physical results. For example, the repeating pattern barely seen in Figure~\ref{Fig_hi_iso2}(i) is typical for cluster spectral function calculations. Also, the momentum resolution is coarse and provides only qualitative result due to a small cluster size and an absence of the intercluster interaction in the computation scheme.

\begin{figure}
\centering
\includegraphics[width=1.0\linewidth]{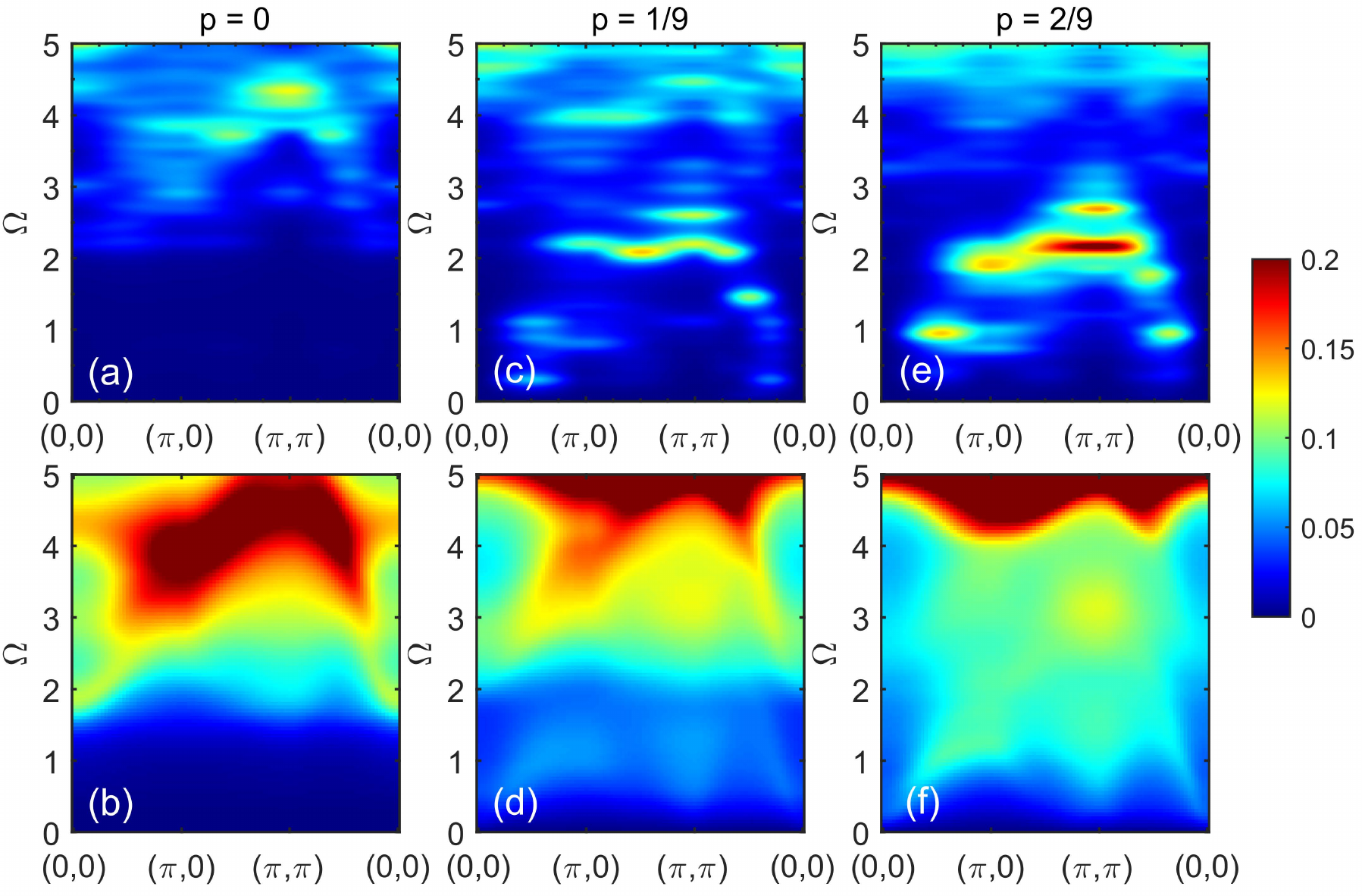}
\caption{\label{Fig_hi_c} Dynamical charge susceptibility $\Pi(\mathbf{q},\Omega)$ within charge-CPT (a),(c),(e) and CPT-RPA (b),(d),(f) for dopings $p=0$ (a),(b), $p=1/9$ (c),(d), and $p=2/9$ (e),(f).}
\end{figure}

Apart from the spin susceptibility, we also calculated the dynamical charge susceptibility. It was obtained in CPT-RPA and in charge-CPT, see Figure~\ref{Fig_hi_c}. Within both approaches, at the increasing doping the spectral weight appears at low energies. In CPT-RPA, the feature reminiscent of noninteracting susceptibility is observed, while in charge-CPT, the spectral weight below 2~eV might resemble some dispersion law with the bandwidth significantly larger than that of the spin excitations. The small cluster size used here does not allow to elucidate complicated charge ordering effects such as those obtained via the quantum Monte Carlo on large clusters~\cite{new_cdw}. The doping evolution of dynamical charge susceptibility within charge-CPT is in agreement with the most general tendencies observed in quantum Monte Carlo~\cite{Grober,Dong19}, but strong size effects in our calculations do not allow us to obtain more subtle features than those discussed above.

\section{\label{Discussion} Discussion}

Now we turn to the qualitative comparison of the obtained results to the experimental spectra from the inelastic neutron scattering (INS) on cuprates. At half filling, cuprates are antiferromagnetic insulators due to a strong Coulomb interaction and their spin spectrum has a spin-wave character~\cite{Coldea01}. Unfortunately, CPT-RPA fails to reproduce a spin-wave spectrum such as observed in cuprates at half filling (the spectrum obtained in this case is uninformative and not shown). On the other hand, spin-CPT shows signatures of such a spectrum and it can be explained as coming from the explicit inclusion of the short-range antiferromagnetic correlations within a cluster.

Addition of charge carriers into the CuO$_2$ planes leads to the formation of a more complicated spectrum, which nature is still under debate~\cite{Tranquada14,Fujita12}. Often, the lower downward dispersing and upper upwards dispersing branches of a hourglass spectrum are discussed. They are reported to have different temperature dependencies~\cite{Sato20} and presumably are of different nature. In the underdoped pseudogap region, several types of behavior were found in different materials. For example, spin spectrum of La$_{2-x}$Sr$_x$CuO$_4$ has an hourglass shape both in superconducting and normal states~\cite{Lipscombe09}, while there are no signatures of the lower downward dispersing branch in the normal state of YBa$_2$Cu$_3$O$_{6+x}$~\cite{Hinkov07}. For underdoped HgBa$_2$CuO$_{4+\delta}$, there is no lower branch neither in the superconducting nor in the normal states~\cite{Chan16}, however, the lower branch was observed near optimal doping in the superconducting state~\cite{Chan16prl}. Both our methods in the underdoped region result in the signatures of the upper branch with the lowest excitation energy at $(\pi,\pi)$ and no presence of low-energy incommensurate excitations. Particularly, the absence of the low-energy incommensurate excitations in CPT-RPA, where the susceptibility is calculated from the electronic structure exhibiting the pseudogap behaviour, might point at the suppression of such excitations by the pseudogap.

In the overdoped regime, both our methods produce the response at lowest energies around the four incommensurate wave vectors located at qualitatively similar positions as in the INS data for overdoped La$_{2-x}$Sr$_x$CuO$_4$~\cite{Wakimoto07,Lipscombe07}. At high energies, CPT-RPA gives a picture somewhat similar to INS, i.e. a weaker broad feature around $(\pi,\pi)$.
In spin-CPT, the response at and close to the $(\pi,\pi)$-point appears only at very high energies, although the general fact of weakening of high-energy response close to the $(\pi,\pi)$-point is similar to what is observed experimentally. Possible explanation of the observed discrepancies is that the RPA-part of the CPT-RPA underestimates the spin-spin correlations in cuprates and thus it can not completely reproduce the spin-wave-like high-energy excitations like those seen in INS. Spin-CPT, in its turn, has coarse momentum and energy resolutions, thus, fails to reproduce fine dispersive features present in CPT-RPA.

RIXS experiments have been successful in studying the high-energy magnetic excitations although with some restrictions on the probed area of wave vectors $\mathbf{q}$. Via RIXS, magnon-like excitations in cuprates were detected dispersing upward to the energies as high as $\sim 400$~meV in the wide doping range from underdoped to heavily overdoped samples~\cite{Letacon11,Letacon13,Wang22,Robarts19}. In all the spin spectra presented here for the wave vectors in the $[0, 0.5\pi]$ range, mainly available in RIXS, an overall magnon-like shape of dispersion can be traced, where the highest response intensity is observed at energies increasing with the increasing wave vector from $(0,0)$ to $(0,\pi)$ and from $(0,0)$ to $(\pi/2,\pi/2)$. The energy width of the response is larger in the first direction than in the second. Momentum-dependent charge excitations were also found in RIXS~\cite{Ishii17}. They are dispersing upward to the energy twice the maximal one of the spin excitations, which is in general consistent with the charge susceptibility spectra obtained here.

\section{\label{Conclusion} Conclusion}

We have formulated a cluster perturbation theory for the two-particle correlation functions. In particular, spin-CPT and charge-CPT for spin and charge dynamical susceptibilities, respectively. Both quantities were calculated by spin(charge)-CPT and CPT-RPA in a wide doping range for the effective two-orbital Hubbard model obtained from the Emery $p-d$ model. In the underdoped case, magnetic response shows signatures of the upper branch of the hourglass dispersion and no presence of the low-energy incommensurate spin excitations. In the overdoped case, both spin-CPT and CPT-RPA produce the low-energy spin excitations located near the four incommensurate wave vectors. These results agree with the spectra for doped cuprates obtained in INS and RIXS. Calculated momentum-dependent charge excitations are in general agreement with RIXS.

One may expect that CPT-RPA overestimates the contribution of the itinerant electrons to a two-particle quantity like the spin susceptibility. But it is important that it bears an effect of the electronic structure that is expected to be obtained quite accurately in CPT. And the electronic structure affects the spin susceptibility through the particle-hole bubble enhanced by the RPA vertex. Spin-CPT, on the other hand, converges to the exact result with the increase of the system size but underestimates the long-range correlations. In general, CPT-based approaches are a simple and economic methods to calculate the momentum- and energy-resolved one- and two-particle correlation functions and allow to take into account short-range correlations exactly. The latter plays a key role in doped high-$T_c$ cuprates.



\authorcontributions{Conceptualization, M.M.K., S.G.O., and V.I.K.; calculations, V.I.K.; formal analysis, S.G.O.; writing, V.I.K., S.V.N., and M.M.K.. All authors have read and agreed to the published version of the manuscript.}

\funding{V.I.K. and S.V.N. acknowledge support by the Krasnoyarsk Regional Science Foundation and Russian Science Foundation grant No. 22-22-20007.}


\conflictsofinterest{The authors declare no conflict of interest.}

\begin{adjustwidth}{-\extralength}{0cm}

\reftitle{References}



\begin{thebibliography}{999}

\bibitem[Le~Tacon et~al.(2011)Le~Tacon, Ghiringhelli, Chaloupka, Sala, Hinkov,
  Haverkort, Minola, Bakr, Zhou, Blanco-Canosa, et~al.]{Letacon11}
Le~Tacon, M.; Ghiringhelli, G.; Chaloupka, J.; Sala, M.M.; Hinkov, V.;
  Haverkort, M.; Minola, M.; Bakr, M.; Zhou, K.; Blanco-Canosa, S.;  et~al.
\newblock Intense paramagnon excitations in a large family of high-temperature
  superconductors.
\newblock {\em Nature Physics} {\bf 2011}, {\em 7},~725--730.
\newblock {\url{https://doi.org/10.1038/nphys2041}}.

\bibitem[Le~Tacon et~al.(2013)Le~Tacon, Minola, Peets, Moretti~Sala,
  Blanco-Canosa, Hinkov, Liang, Bonn, Hardy, Lin, Schmitt, Braicovich,
  Ghiringhelli, and Keimer]{Letacon13}
Le~Tacon, M.; Minola, M.; Peets, D.C.; Moretti~Sala, M.; Blanco-Canosa, S.;
  Hinkov, V.; Liang, R.; Bonn, D.A.; Hardy, W.N.; Lin, C.T.;  et~al.
\newblock Dispersive spin excitations in highly overdoped cuprates revealed by
  resonant inelastic x-ray scattering.
\newblock {\em Phys. Rev. B} {\bf 2013}, {\em 88},~020501.
\newblock {\url{https://doi.org/10.1103/PhysRevB.88.020501}}.

\bibitem[Wang et~al.(2022)Wang, He, Yang, Garcia-Fernandez, Nag, Zhou, Minola,
  Tacon, Keimer, Peng, et~al.]{Wang22}
Wang, L.; He, G.; Yang, Z.; Garcia-Fernandez, M.; Nag, A.; Zhou, K.; Minola,
  M.; Tacon, M.L.; Keimer, B.; Peng, Y.;  et~al.
\newblock Paramagnons and high-temperature superconductivity in a model family
  of cuprates.
\newblock {\em Nature Communications} {\bf 2022}, {\em 13},~3163.
\newblock {\url{https://doi.org/10.1038/s41467-022-30918-z}}.

\bibitem[Robarts et~al.(2019)Robarts, Barth\'elemy, Kummer,
  Garc\'{\i}a-Fern\'andez, Li, Nag, Walters, Zhou, and Hayden]{Robarts19}
Robarts, H.C.; Barth\'elemy, M.; Kummer, K.; Garc\'{\i}a-Fern\'andez, M.; Li,
  J.; Nag, A.; Walters, A.C.; Zhou, K.J.; Hayden, S.M.
\newblock Anisotropic damping and wave vector dependent susceptibility of the
  spin fluctuations in
  ${\mathrm{La}}_{2\ensuremath{-}x}{\mathrm{Sr}}_{x}{\mathrm{CuO}}_{4}$ studied
  by resonant inelastic x-ray scattering.
\newblock {\em Phys. Rev. B} {\bf 2019}, {\em 100},~214510.
\newblock {\url{https://doi.org/10.1103/PhysRevB.100.214510}}.

\bibitem[Ishii et~al.(2017)Ishii, Tohyama, Asano, Sato, Fujita, Wakimoto,
  Tustsui, Sota, Miyawaki, Niwa, Harada, Pelliciari, Huang, Schmitt, Yamamoto,
  and Mizuki]{Ishii17}
Ishii, K.; Tohyama, T.; Asano, S.; Sato, K.; Fujita, M.; Wakimoto, S.; Tustsui,
  K.; Sota, S.; Miyawaki, J.; Niwa, H.;  et~al.
\newblock Observation of momentum-dependent charge excitations in hole-doped
  cuprates using resonant inelastic x-ray scattering at the oxygen $K$ edge.
\newblock {\em Phys. Rev. B} {\bf 2017}, {\em 96},~115148.
\newblock {\url{https://doi.org/10.1103/PhysRevB.96.115148}}.

\bibitem[Sobota et~al.(2021)Sobota, He, and Shen]{Sobota_2021}
Sobota, J.A.; He, Y.; Shen, Z.X.
\newblock Angle-resolved photoemission studies of quantum materials.
\newblock {\em Rev. Mod. Phys.} {\bf 2021}, {\em 93},~025006.
\newblock {\url{https://doi.org/10.1103/RevModPhys.93.025006}}.

\bibitem[Tranquada et~al.(2014)Tranquada, Xu, and Zaliznyak]{Tranquada14}
Tranquada, J.M.; Xu, G.; Zaliznyak, I.A.
\newblock Superconductivity, antiferromagnetism, and neutron scattering.
\newblock {\em Journal of Magnetism and Magnetic Materials} {\bf 2014}, {\em
  350},~148--160.
\newblock {\url{https://doi.org/https://doi.org/10.1016/j.jmmm.2013.09.029}}.

\bibitem[Fujita et~al.(2012)Fujita, Hiraka, Matsuda, Matsuura, M.~Tranquada,
  Wakimoto, Xu, and Yamada]{Fujita12}
Fujita, M.; Hiraka, H.; Matsuda, M.; Matsuura, M.; M.~Tranquada, J.; Wakimoto,
  S.; Xu, G.; Yamada, K.
\newblock Progress in Neutron Scattering Studies of Spin Excitations in High-Tc
  Cuprates.
\newblock {\em Journal of the Physical Society of Japan} {\bf 2012}, {\em
  81},~011007.
\newblock {\url{https://doi.org/10.1143/JPSJ.81.011007}}.

\bibitem[Sato et~al.(2020)Sato, Ikeuchi, Kajimoto, Wakimoto, Arai, and
  Fujita]{Sato20}
Sato, K.; Ikeuchi, K.; Kajimoto, R.; Wakimoto, S.; Arai, M.; Fujita, M.
\newblock Coexistence of Two Components in Magnetic Excitations of La2-xSrxCuO4
  (x = 0.10 and 0.16).
\newblock {\em Journal of the Physical Society of Japan} {\bf 2020}, {\em
  89},~114703.
\newblock {\url{https://doi.org/10.7566/JPSJ.89.114703}}.

\bibitem[Lipscombe et~al.(2009)Lipscombe, Vignolle, Perring, Frost, and
  Hayden]{Lipscombe09}
Lipscombe, O.J.; Vignolle, B.; Perring, T.G.; Frost, C.D.; Hayden, S.M.
\newblock Emergence of Coherent Magnetic Excitations in the High Temperature
  Underdoped
  ${\mathrm{La}}_{2\ensuremath{-}x}{\mathrm{Sr}}_{x}{\mathrm{CuO}}_{4}$
  Superconductor at Low Temperatures.
\newblock {\em Phys. Rev. Lett.} {\bf 2009}, {\em 102},~167002.
\newblock {\url{https://doi.org/10.1103/PhysRevLett.102.167002}}.

\bibitem[Hinkov et~al.(2007)Hinkov, Bourges, Pailhes, Sidis, Ivanov, Frost,
  Perring, Lin, Chen, and Keimer]{Hinkov07}
Hinkov, V.; Bourges, P.; Pailhes, S.; Sidis, Y.; Ivanov, A.; Frost, C.;
  Perring, T.; Lin, C.; Chen, D.; Keimer, B.
\newblock Spin dynamics in the pseudogap state of a high-temperature
  superconductor.
\newblock {\em Nature Physics} {\bf 2007}, {\em 3},~780--785.
\newblock {\url{https://doi.org/10.1038/nphys720}}.

\bibitem[Chan et~al.(2016{\natexlab{a}})Chan, Dorow, Mangin-Thro, Tang, Ge,
  Veit, Yu, Zhao, Christianson, Park, et~al.]{Chan16}
Chan, M.; Dorow, C.; Mangin-Thro, L.; Tang, Y.; Ge, Y.; Veit, M.; Yu, G.; Zhao,
  X.; Christianson, A.; Park, J.;  et~al.
\newblock Commensurate antiferromagnetic excitations as a signature of the
  pseudogap in the tetragonal high-T c cuprate HgBa2CuO4+ $\delta$.
\newblock {\em Nature communications} {\bf 2016}, {\em 7},~10819.
\newblock {\url{https://doi.org/10.1038/ncomms10819}}.

\bibitem[Chan et~al.(2016{\natexlab{b}})Chan, Tang, Dorow, Jeong, Mangin-Thro,
  Veit, Ge, Abernathy, Sidis, Bourges, and Greven]{Chan16prl}
Chan, M.K.; Tang, Y.; Dorow, C.J.; Jeong, J.; Mangin-Thro, L.; Veit, M.J.; Ge,
  Y.; Abernathy, D.L.; Sidis, Y.; Bourges, P.;  et~al.
\newblock Hourglass Dispersion and Resonance of Magnetic Excitations in the
  Superconducting State of the Single-Layer Cuprate
  ${\mathrm{HgBa}}_{2}{\mathrm{CuO}}_{4+\ensuremath{\delta}}$ Near Optimal
  Doping.
\newblock {\em Phys. Rev. Lett.} {\bf 2016}, {\em 117},~277002.
\newblock {\url{https://doi.org/10.1103/PhysRevLett.117.277002}}.

\bibitem[S\'en\'echal et~al.(2000)S\'en\'echal, Perez, and
  Pioro-Ladri\`ere]{Senechal00}
S\'en\'echal, D.; Perez, D.; Pioro-Ladri\`ere, M.
\newblock Spectral Weight of the Hubbard Model through Cluster Perturbation
  Theory.
\newblock {\em Phys. Rev. Lett.} {\bf 2000}, {\em 84},~522--525.
\newblock {\url{https://doi.org/10.1103/PhysRevLett.84.522}}.

\bibitem[S\'en\'echal et~al.(2002)S\'en\'echal, Perez, and Plouffe]{Senechal02}
S\'en\'echal, D.; Perez, D.; Plouffe, D.
\newblock Cluster perturbation theory for Hubbard models.
\newblock {\em Phys. Rev. B} {\bf 2002}, {\em 66},~075129.
\newblock {\url{https://doi.org/10.1103/PhysRevB.66.075129}}.

\bibitem[Maier et~al.(2005)Maier, Jarrell, Pruschke, and Hettler]{Maier2005}
Maier, T.; Jarrell, M.; Pruschke, T.; Hettler, M.H.
\newblock Quantum cluster theories.
\newblock {\em Rev. Mod. Phys.} {\bf 2005}, {\em 77},~1027--1080.
\newblock {\url{https://doi.org/10.1103/RevModPhys.77.1027}}.

\bibitem[Brehm et~al.(2010)Brehm, Arrigoni, Aichhorn, and Hanke]{Brehm_2010}
Brehm, S.; Arrigoni, E.; Aichhorn, M.; Hanke, W.
\newblock Theory of two-particle excitations and the magnetic susceptibility in
  high-Tc cuprate superconductors.
\newblock {\em Europhysics Letters} {\bf 2010}, {\em 89},~27005.
\newblock {\url{https://doi.org/10.1209/0295-5075/89/27005}}.

\bibitem[Kung et~al.(2017)Kung, Bazin, Wohlfeld, Wang, Chen, Jia, Johnston,
  Moritz, Mila, and Devereaux]{Kung17}
Kung, Y.F.; Bazin, C.; Wohlfeld, K.; Wang, Y.; Chen, C.C.; Jia, C.J.; Johnston,
  S.; Moritz, B.; Mila, F.; Devereaux, T.P.
\newblock Numerically exploring the 1D-2D dimensional crossover on spin
  dynamics in the doped Hubbard model.
\newblock {\em Phys. Rev. B} {\bf 2017}, {\em 96},~195106.
\newblock {\url{https://doi.org/10.1103/PhysRevB.96.195106}}.

\bibitem[Raum et~al.(2020)Raum, Alvarez, Maier, and Scarola]{Raum20}
Raum, P.T.; Alvarez, G.; Maier, T.; Scarola, V.W.
\newblock Two-particle correlation functions in cluster perturbation theory:
  Hubbard spin susceptibilities.
\newblock {\em Phys. Rev. B} {\bf 2020}, {\em 101},~075122.
\newblock {\url{https://doi.org/10.1103/PhysRevB.101.075122}}.

\bibitem[Emery(1987)]{Emery}
Emery, V.J.
\newblock Theory of high-${\mathrm{T}}_{\mathrm{c}}$ superconductivity in
  oxides.
\newblock {\em Phys. Rev. Lett.} {\bf 1987}, {\em 58},~2794--2797.
\newblock {\url{https://doi.org/10.1103/PhysRevLett.58.2794}}.

\bibitem[Chen et~al.(2015)Chen, van Veenendaal, Devereaux, and
  Wohlfeld]{Chen15}
Chen, C.C.; van Veenendaal, M.; Devereaux, T.P.; Wohlfeld, K.
\newblock Fractionalization, entanglement, and separation: Understanding the
  collective excitations in a spin-orbital chain.
\newblock {\em Phys. Rev. B} {\bf 2015}, {\em 91},~165102.
\newblock {\url{https://doi.org/10.1103/PhysRevB.91.165102}}.

\bibitem[P\"arschke et~al.(2019)P\"arschke, Wang, Moritz, Devereaux, Chen, and
  Wohlfeld]{Parschke19}
P\"arschke, E.M.; Wang, Y.; Moritz, B.; Devereaux, T.P.; Chen, C.C.; Wohlfeld,
  K.
\newblock Numerical investigation of spin excitations in a doped spin chain.
\newblock {\em Phys. Rev. B} {\bf 2019}, {\em 99},~205102.
\newblock {\url{https://doi.org/10.1103/PhysRevB.99.205102}}.

\bibitem[Nikolaev and Korshunov(2016)]{Nikolaev16}
Nikolaev, S.V.; Korshunov, M.M.
\newblock Spin and Charge Susceptibilities of the Two-Orbital Model within the
  Cluster Perturbation Theory for Fe-Based Materials.
\newblock {\em J. Supercond. Nov. Magn.} {\bf 2016}, {\em 29},~3093.
\newblock {\url{https://doi.org/10.1007/s10948-016-3784-8}}.

\bibitem[Korshunov et~al.(2005)Korshunov, Gavrichkov, Ovchinnikov, Nekrasov,
  Pchelkina, and Anisimov]{Korshunov05}
Korshunov, M.M.; Gavrichkov, V.A.; Ovchinnikov, S.G.; Nekrasov, I.A.;
  Pchelkina, Z.V.; Anisimov, V.I.
\newblock Hybrid LDA and generalized tight-binding method for electronic
  structure calculations of strongly correlated electron systems.
\newblock {\em Phys. Rev. B} {\bf 2005}, {\em 72},~165104.
\newblock {\url{https://doi.org/10.1103/PhysRevB.72.165104}}.

\bibitem[Anisimov et~al.(2005)Anisimov, Kondakov, Kozhevnikov, Nekrasov,
  Pchelkina, Allen, Mo, Kim, Metcalf, Suga, Sekiyama, Keller, Leonov, Ren, and
  Vollhardt]{Anisimov05}
Anisimov, V.I.; Kondakov, D.E.; Kozhevnikov, A.V.; Nekrasov, I.A.; Pchelkina,
  Z.V.; Allen, J.W.; Mo, S.K.; Kim, H.D.; Metcalf, P.; Suga, S.;  et~al.
\newblock Full orbital calculation scheme for materials with strongly
  correlated electrons.
\newblock {\em Phys. Rev. B} {\bf 2005}, {\em 71},~125119.
\newblock {\url{https://doi.org/10.1103/PhysRevB.71.125119}}.

\bibitem[Hybertsen et~al.(1989)Hybertsen, Schl\"uter, and
  Christensen]{Hybertsen89}
Hybertsen, M.S.; Schl\"uter, M.; Christensen, N.E.
\newblock Calculation of Coulomb-interaction parameters for
  ${\mathrm{La}}_{2}$${\mathrm{CuO}}_{4}$ using a
  constrained-density-functional approach.
\newblock {\em Phys. Rev. B} {\bf 1989}, {\em 39},~9028--9041.
\newblock {\url{https://doi.org/10.1103/PhysRevB.39.9028}}.

\bibitem[Shastry(1989)]{Shastry}
Shastry, B.S.
\newblock t-J model and nuclear magnetic relaxation in
  high-${\mathrm{T}}_{\mathrm{c}}$ materials.
\newblock {\em Phys. Rev. Lett.} {\bf 1989}, {\em 63},~1288--1291.
\newblock {\url{https://doi.org/10.1103/PhysRevLett.63.1288}}.

\bibitem[Lovtsov and Yushankhai(1991)]{Lovtsov91}
Lovtsov, S.; Yushankhai, V.
\newblock Effective singlet-triplet model for CuO2 plane in oxide
  superconductors: the change fluctuation regime.
\newblock {\em Physica C: Superconductivity} {\bf 1991}, {\em 179},~159 -- 166.
\newblock {\url{https://doi.org/https://doi.org/10.1016/0921-4534(91)90024-S}}.

\bibitem[Jefferson et~al.(1992)Jefferson, Eskes, and Feiner]{Jefferson92}
Jefferson, J.H.; Eskes, H.; Feiner, L.F.
\newblock Derivation of a single-band model for ${\mathrm{CuO}}_{2}$ planes by
  a cell-perturbation method.
\newblock {\em Phys. Rev. B} {\bf 1992}, {\em 45},~7959--7972.
\newblock {\url{https://doi.org/10.1103/PhysRevB.45.7959}}.

\bibitem[Sch\"uttler and Fedro(1992)]{Schuttler92}
Sch\"uttler, H.B.; Fedro, A.J.
\newblock Copper-oxygen charge excitations and the effective-single-band theory
  of cuprate superconductors.
\newblock {\em Phys. Rev. B} {\bf 1992}, {\em 45},~7588--7591.
\newblock {\url{https://doi.org/10.1103/PhysRevB.45.7588}}.

\bibitem[Feiner et~al.(1996)Feiner, Jefferson, and Raimondi]{Feiner96}
Feiner, L.F.; Jefferson, J.H.; Raimondi, R.
\newblock Effective single-band models for the high-${\mathit{T}}_{\mathit{c}}$
  cuprates. I. Coulomb interactions.
\newblock {\em Phys. Rev. B} {\bf 1996}, {\em 53},~8751--8773.
\newblock {\url{https://doi.org/10.1103/PhysRevB.53.8751}}.

\bibitem[Raimondi et~al.(1996)Raimondi, Jefferson, and Feiner]{Raimondi96}
Raimondi, R.; Jefferson, J.H.; Feiner, L.F.
\newblock Effective single-band models for the high-${\mathit{T}}_{\mathit{c}}$
  cuprates. II. Role of apical oxygen.
\newblock {\em Phys. Rev. B} {\bf 1996}, {\em 53},~8774--8788.
\newblock {\url{https://doi.org/10.1103/PhysRevB.53.8774}}.

\bibitem[Gavrichkov et~al.(2000)Gavrichkov, Ovchinnikov, Borisov, and
  Goryachev]{Gavrichkov00}
Gavrichkov, V.A.; Ovchinnikov, S.G.; Borisov, A.A.; Goryachev, E.G.
\newblock Evolution of the band structure of quasiparticles with doping in
  copper oxides on the basis of a generalized tight-binding method.
\newblock {\em Journal of Experimental and Theoretical Physics} {\bf 2000},
  {\em 91},~369--383.
\newblock {\url{https://doi.org/10.1134/1.1311997}}.

\bibitem[Makarov et~al.(2015)Makarov, Shneyder, Kozlov, and
  Ovchinnikov]{Makarov15}
Makarov, I.A.; Shneyder, E.I.; Kozlov, P.A.; Ovchinnikov, S.G.
\newblock Polaronic approach to strongly correlated electron systems with
  strong electron-phonon interaction.
\newblock {\em Phys. Rev. B} {\bf 2015}, {\em 92},~155143.
\newblock {\url{https://doi.org/10.1103/PhysRevB.92.155143}}.

\bibitem[Shneyder et~al.(2020)Shneyder, Nikolaev, Zotova, Kaldin, and
  Ovchinnikov]{Shneyder20}
Shneyder, E.I.; Nikolaev, S.V.; Zotova, M.V.; Kaldin, R.A.; Ovchinnikov, S.G.
\newblock Polaron transformations in the realistic model of the strongly
  correlated electron system.
\newblock {\em Phys. Rev. B} {\bf 2020}, {\em 101},~235114.
\newblock {\url{https://doi.org/10.1103/PhysRevB.101.235114}}.

\bibitem[Kung et~al.(2016)Kung, Chen, Wang, Huang, Nowadnick, Moritz,
  Scalettar, Johnston, and Devereaux]{Kung16}
Kung, Y.F.; Chen, C.C.; Wang, Y.; Huang, E.W.; Nowadnick, E.A.; Moritz, B.;
  Scalettar, R.T.; Johnston, S.; Devereaux, T.P.
\newblock Characterizing the three-orbital Hubbard model with determinant
  quantum Monte Carlo.
\newblock {\em Phys. Rev. B} {\bf 2016}, {\em 93},~155166.
\newblock {\url{https://doi.org/10.1103/PhysRevB.93.155166}}.

\bibitem[Huang et~al.(2022)Huang, Ding, Liu, and Wang]{Huang21}
Huang, E.W.; Ding, S.; Liu, J.; Wang, Y.
\newblock Determinantal quantum Monte Carlo solver for cluster perturbation
  theory.
\newblock {\em Phys. Rev. Res.} {\bf 2022}, {\em 4},~L042015.
\newblock {\url{https://doi.org/10.1103/PhysRevResearch.4.L042015}}.

\bibitem[Graser et~al.(2009)Graser, Maier, Hirschfeld, and Scalapino]{Graser09}
Graser, S.; Maier, T.A.; Hirschfeld, P.J.; Scalapino, D.J.
\newblock Near-degeneracy of several pairing channels in multiorbital models
  for the Fe pnictides.
\newblock {\em New Journal of Physics} {\bf 2009}, {\em 11},~025016.
\newblock {\url{https://doi.org/10.1088/1367-2630/11/2/025016}}.

\bibitem[Korshunov(2018)]{Korshunov_chapter}
Korshunov, M.M.
\newblock Itinerant spin fluctuations in iron-based superconductors. In {\em
  Perturbation Theory: Advances in Research and Applications}; Pirogov, Z.,
  Ed.; Nova Science Publishers Inc.: New York,  2018; chapter~2, pp. 61--138.

\bibitem[Nikolaev and Ovchinnikov(2010)]{Nikolaev10}
Nikolaev, S.V.; Ovchinnikov, S.G.
\newblock Cluster perturbation theory in Hubbard model exactly taking into
  account the short-range magnetic order in 2 {\texttimes} 2 cluster.
\newblock {\em JETP} {\bf 2010}, {\em 111},~635--644.
\newblock {\url{https://doi.org/10.1134/S1063776110100146}}.

\bibitem[Nikolaev and Ovchinnikov(2012)]{Nikolaev12}
Nikolaev, S.V.; Ovchinnikov, S.G.
\newblock Effect of hole doping on the electronic structure and the Fermi
  surface in the Hubbard model within norm-conserving cluster pertubation
  theory.
\newblock {\em Journal of Experimental and Theoretical Physics} {\bf 2012},
  {\em 114},~118--131.
\newblock {\url{https://doi.org/10.1134/S1063776111150143}}.

\bibitem[Kuz'min et~al.(2014)Kuz'min, Nikolaev, and Ovchinnikov]{Kuzmin14}
Kuz'min, V.I.; Nikolaev, S.V.; Ovchinnikov, S.G.
\newblock Comparison of the electronic structure of the Hubbard and
  $t\ensuremath{-}J$ models within the cluster perturbation theory.
\newblock {\em Phys. Rev. B} {\bf 2014}, {\em 90},~245104.
\newblock {\url{https://doi.org/10.1103/PhysRevB.90.245104}}.

\bibitem[Kuz'min et~al.(2020)Kuz'min, Visotin, Nikolaev, and
  Ovchinnikov]{Kuzmin20}
Kuz'min, V.I.; Visotin, M.A.; Nikolaev, S.V.; Ovchinnikov, S.G.
\newblock Doping and temperature evolution of pseudogap and spin-spin
  correlations in the two-dimensional Hubbard model.
\newblock {\em Phys. Rev. B} {\bf 2020}, {\em 101},~115141.
\newblock {\url{https://doi.org/10.1103/PhysRevB.101.115141}}.

\bibitem[Mai et~al.(2022)Mai, Nichols, Karakuzu, Bao, Del~Maestro, Maier, and
  Johnston]{new_cdw}
Mai, P.; Nichols, N.S.; Karakuzu, S.; Bao, F.; Del~Maestro, A.; Maier, T.A.;
  Johnston, S.
\newblock Robust charge-density wave correlations in the electron-doped
  single-band Hubbard model,  2022.
\newblock {\url{https://doi.org/10.48550/ARXIV.2210.14930}}.

\bibitem[Gr\"ober et~al.(2000)Gr\"ober, Eder, and Hanke]{Grober}
Gr\"ober, C.; Eder, R.; Hanke, W.
\newblock Anomalous low-doping phase of the Hubbard model.
\newblock {\em Phys. Rev. B} {\bf 2000}, {\em 62},~4336--4352.
\newblock {\url{https://doi.org/10.1103/PhysRevB.62.4336}}.

\bibitem[Dong et~al.(2019)Dong, Chen, and Gull]{Dong19}
Dong, X.; Chen, X.; Gull, E.
\newblock Dynamical charge susceptibility in the Hubbard model.
\newblock {\em Phys. Rev. B} {\bf 2019}, {\em 100},~235107.
\newblock {\url{https://doi.org/10.1103/PhysRevB.100.235107}}.

\bibitem[Coldea et~al.(2001)Coldea, Hayden, Aeppli, Perring, Frost, Mason,
  Cheong, and Fisk]{Coldea01}
Coldea, R.; Hayden, S.M.; Aeppli, G.; Perring, T.G.; Frost, C.D.; Mason, T.E.;
  Cheong, S.W.; Fisk, Z.
\newblock Spin Waves and Electronic Interactions in
  ${\mathrm{La}}_{2}{\mathrm{CuO}}_{4}$.
\newblock {\em Phys. Rev. Lett.} {\bf 2001}, {\em 86},~5377--5380.
\newblock {\url{https://doi.org/10.1103/PhysRevLett.86.5377}}.

\bibitem[Wakimoto et~al.(2007)Wakimoto, Yamada, Tranquada, Frost, Birgeneau,
  and Zhang]{Wakimoto07}
Wakimoto, S.; Yamada, K.; Tranquada, J.M.; Frost, C.D.; Birgeneau, R.J.; Zhang,
  H.
\newblock Disappearance of Antiferromagnetic Spin Excitations in Overdoped
  ${\mathrm{La}}_{2\ensuremath{-}x}{\mathrm{Sr}}_{x}{\mathrm{CuO}}_{4}$.
\newblock {\em Phys. Rev. Lett.} {\bf 2007}, {\em 98},~247003.
\newblock {\url{https://doi.org/10.1103/PhysRevLett.98.247003}}.

\bibitem[Lipscombe et~al.(2007)Lipscombe, Hayden, Vignolle, McMorrow, and
  Perring]{Lipscombe07}
Lipscombe, O.J.; Hayden, S.M.; Vignolle, B.; McMorrow, D.F.; Perring, T.G.
\newblock Persistence of High-Frequency Spin Fluctuations in Overdoped
  Superconducting
  ${\mathrm{La}}_{2\ensuremath{-}x}{\mathrm{Sr}}_{x}{\mathrm{CuO}}_{4}$
  ($x=0.22$).
\newblock {\em Phys. Rev. Lett.} {\bf 2007}, {\em 99},~067002.
\newblock {\url{https://doi.org/10.1103/PhysRevLett.99.067002}}.

\end{thebibliography}




\end{adjustwidth}
\end{document}